\documentclass[
 manuscript,
 ]{aastex63}
\usepackage{amsmath}
\usepackage{float}

\received{}
\revised{}
\accepted{}
\submitjournal{\apj}

\shorttitle{KAWs in the magnetosphere}
\shortauthors{Moya et al.}

\begin{document}

\title{The role of O+ and He+ on the propagation of Kinetic Alfvén Waves in the Earth’s inner magnetosphere}

\correspondingauthor{Pablo S. Moya}
\email{pablo.moya@uchile.cl}

\author[0000-0002-9161-0888]{Pablo S. Moya}
\affiliation{Departamento de F\'isica, Facultad de Ciencias, Universidad de Chile, Santiago, Chile}

\author[0000-0003-2430-6058]{Bea Zenteno-Quinteros}
\affiliation{Departamento de F\'isica, Facultad de Ciencias, Universidad de Chile, Santiago, Chile}

\author[0000-0003-4988-4348]{Iv\'an Gallo-M\'endez}
\affiliation{Departamento de F\'isica, Facultad de Ciencias, Universidad de Chile, Santiago, Chile}

\author[0000-0003-1210-167X]{V\'ictor A. Pinto}
\affiliation{Departamento de F\'isica, Universidad de Santiago de Chile, Santiago, Chile}
\affiliation{Institute for the Study of Earth, Oceans, and Space, University of New Hampshire Durham, NH, USA}


\begin{abstract}
Interactions between plasma particles and electromagnetic waves play a crucial role in the dynamics and regulation of the state of space environments. From plasma physics theory, the characteristics of the waves and their interactions with the plasma strongly depend on the composition of the plasma, among other factors. In the case of the Earth's magnetosphere, the plasma is usually composed by electrons, protons, O+, and He+ ions, all with their particular properties and characteristics. Here, using plasma parameters relevant for the inner magnetosphere we study the dispersion properties of Kinetic Alfvén Waves (KAW) in a plasma composed by electrons, protons, He+ ions, and O+ ions. We show that heavy ions induce significant changes to the dispersion properties of KAW, such as polarization, compressibility, and electric-to-magnetic amplitude ratio, and therefore the propagation of Kinetic Alfvén Waves is highly determined by the relative abundance of He+ and O+ in the plasma. These results when discussed in the context of observations in the Earth's magnetosphere, suggest that for many types of studies based on theory and numerical simulations, the inclusion of heavy ions should be customary for realistic modeling of plasma phenomena in the inner magnetosphere or other space environments in which heavy ions can contribute with a substantial portion of the plasma such as planetary magnetospheres and comet plasma tails.
\end{abstract}

%\onecolumn \maketitle %\normalsize \vfill

\section{Introduction}
\label{ref:intro}

First proposed by~\citet{hasegawa1976}, Kinetic Alfvén Waves (KAWs) appear at quasi-perpendicular wave-normal angles, when kinetic effects of electrons become relevant. For that to occur, KAWs require the ion gyroradius $\rho_i$ to be similar to the perpendicular wavelength ($k_{\perp} \rho_i \sim 1$) with $k_\perp$ the wave-number in the direction perpendicular to the background magnetic field~\citep{Narita2020}. The gyroradius is given by $\rho_i = \alpha_i/\Omega_i$, where $\alpha_i$ and $\Omega_i$ are the thermal speed and gyrofrequency of the ions. Kinetic Alfvén Waves are mainly characterized for their right-handed polarization in the plasma frame~\citep{gary1986} in contrast to the left-handed polarized Electromagnetic Ion-Cyclotron (EMIC) waves. Therefore, the wave-normal angle in which Alfvénic waves shift from EMIC waves to KAWs strongly depends on the plasma beta parameter~\citep{gary1993,Gary2004}. 

KAW-particle interactions are mostly resonant with electrons and non-resonant with ions~\citep{Lakhina2008,Barik2020}, and as such, have been proposed as a possible mechanism for the energy transfer from ion to electron scales~\citep{nandal2016}. Thus, the interest to study the role of these waves in space plasmas such as the Earth's magnetospheric environment has been increasing in the last few decades. KAWs have been observed to play important roles in several processes such as energy dissipation at kinetic turbulence~\citep{Chasapis2018,Macek2018,Dwivedi2019} and reconnection in the magnetosheath~\citep{chaston2005,Boldyrev2019}; electron acceleration in the plasma sheet~\citep{wygant2002,Duan2012}; energization of the ionosphere~\citep{Keiling2019,Cheng2020} or ionospheric outflow~\citep{Chaston2016}. These waves have also been observed in the inner magnetosphere during geomagnetic storms~\citep{chaston2014,moya2015}. 

Besides their right-hand polarization, in the plasma frame KAWs exhibit positive magnetic helicity and other signature dispersion properties that are helpful to identify these waves using observations. For example, due to their quasi-perpendicular wave-normal angles, KAWs usually have large magnetic compressibility, and have larger parallel fluctuating electric fields when compared to Alfvénic EMIC waves~\citep{gary1986,voitenko2006,chaston2014,moya2015}. Moreover, the $\delta E_\parallel/\delta B_\bot$ spectral ratio between field-aligned electric fluctuations, $\delta E_\parallel$, and transverse magnetic ﬁeld ﬂuctuations, $\delta B_\bot$, is an important signature to distinguish between EMIC waves, magnetosonic waves (MSW) and KAWs. In the case of EMIC waves, $\delta E_\parallel/\delta B_\bot$ have values on the order of the local Alfvén speed, $V_A = B_0/\sqrt{4\pi n_0 m_p}$ (where $m_p$ is the proton mass and $n_0$ is the total number density), whereas for KAW this ratio is several times $V_A$ and increases with increasing frequency~\citep{chaston2014,moya2015}. The $\delta E_\parallel/\delta B_\bot$ spectrum is also useful to distinguish between MSW and KAW modes, as the magnetosonic wave-mode has a monotonically increasing $\delta E_\parallel/\delta B_\bot$ ratio frequency spectrum, whereas KAW have a $\delta E_\parallel/\delta B_\bot$ ratio spectrum with a local maximum at a given frequency and then decreases for larger frequencies~\citep{salem2012,moya2015}. In summary, all these dispersion properties provide useful information to identify and distinguish different Alfvénic modes using in situ space plasma waves observations~\citep{Narita2020}.

From the theoretical point of view, most of the mentioned dispersion properties of KAWs have in general been studied in proton-electron plasmas or electron-ion plasmas. However, most astrophysical and space plasmas are multi-species. Thus, in these systems it is expected that the dispersion properties of EMIC or KAW modes to be dependent not only on the parameters of proton and electron distributions, but also on the plasma composition and the parameters of heavier ions such as O+ ions~\citep{moya2021}, O+ and He+ ions~\citep{moya2015}, or N+ ions \citep{Bashir2018}. The impact of heavier ions may be especially relevant in the inner magnetosphere during geomagnetic storms as there is a strong dependence between the relative abundance of protons, O+ ions and He+ ions and geomagnetic activity. It has been shown that due to ionospheric outflow during geomagnetic storms, O+ ions can dominate the plasma composition in the ring current region~\citep{daglis1997,daglis1999,Yue2018}, and that the abundance of each ion species is highly dependent on the radial distance to the Earth ($L$-shell) and the strength of the geomagnetic storm~\citep{jahn2017}. Thus, depending on the level of geomagnetic activity, the inner magnetosphere composition may favor or hinder the propagation of KAWs, and therefore the kinetic processes mediated by these waves. In this context, during the recent years, studies have shown that ion beams can provide the free energy necessary to generate KAWs in the inner magnetosphere and auroral region~\citep{Lakhina2008,Barik2019a,Barik2019b, Barik2019c,Barik2021}. In addition,~\citet{moya2021} computed the exact numerical dispersion relation of KAWs, without the use of approximations, considering different abundances of O+, using plasma parameters relevant to the Earth inner magnetosphere and cometary environments. They found that the presence of O+ ions allows the propagation of weakly damped or unstable KAW modes at smaller wave-normal angles than if the plasma were composed only by protons and electrons. However, isotropic heavy ions can drastically reduce the growth rates of unstable KAWs triggered by anisotropic protons, or even inhibit the instability if their relative abundance is large enough. Therefore, heavy ions may play an important role mediating energy transfer processes from large to small scales through wave–particle interactions between plasma particle and KAWs.

In this work we study the effect of magnetospheric heavy ions such as O+ and He+, in combination with protons and electrons, on the dispersion relation and dispersion properties of KAWs and compare them with the pure electron-proton case, using plasma parameters traditionally observed during geomagnetic storms in the Earth's inner magnetosphere. In section \ref{ref:linear} we present the Vlasov dispersion relation and dispersion properties of KAW in a multi-species warm plasma (a plasma where each species has a not zero temperature) in which each species follows a bi-Maxwellian velocity distribution function. In section~\ref{ref:results}, considering plasma parameters relevant to the inner magnetosphere, we compute the exact numerical dispersion relation and spectral properties for Alfvénic waves at different propagation angle, comparing the results obtained in a simple electron-proton plasma, and the results when magetospheric heavy ions (He+ and O+) are considered. Finally, in section~\ref{ref:conclusions} we summarize our numerical results, outline the main conclusions, and discuss the physical scope and relevance of our findings.  

\section{Linear analysis: Vlasov-Maxwell dispersion relation and kinetic dispersion properties}
\label{ref:linear}

Let's consider a warm multi-species plasma in the presence of a background magnetic field $\mathbf{B}_0 = B_0\,\hat z$, in which each species $s$ follows a bi-Maxwellian velocity distribution function (VDF): 
\begin{equation}
\label{eq:vdf}
    f_s(v_\bot, v_\parallel)=\dfrac{n_{0 s}}{\pi^{3/2}\alpha^2_{\bot s}\alpha_{\parallel s}}\exp\left(-\dfrac{v^2_\perp}{\alpha^2_{\bot s}}-\dfrac{v_\parallel^2}{\alpha^2_{\parallel s}}\right)\,,
\end{equation}
where $\alpha^2_{\bot s} = 2k_BT_\bot/m_s$ and $\alpha^2_{\parallel s} = 2k_BT_\parallel/m_s$ are the squares of the thermal speeds of the species $s$, and $T_{\bot s}$ and $T_{\parallel s}$ are the perpendicular and parallel temperatures with respect to $\mathbf{B}_0$. Also, in Equation (\ref{eq:vdf}), $n_{0s}$, and $m_s$ correspond to the equilibrium number density and mass of the $s$-th species, respectively, and $k_B$ is the Boltzmann constant. In such plasma, the warm Vlasov-Maxwell kinetic dispersion relation for electromagnetic waves is determined by
\begin{equation}
\label{eq:dr}
    D(\omega,\,\mathbf{k}; \rm{pp})\,\delta\mathbf{E_k}(\omega, \mathbf{k}) = 0\,
\end{equation}
or $|D(\omega,\,\mathbf{k};\rm{pp})| = 0$, where the so-called dispersion tensor $D$ represents the linear response of the plasma media to electromagnetic perturbations, and is a function of the frequency $\omega$, wave-vector $\mathbf{k}$, and the macroscopic plasma parameters here denoted by \textrm{pp}. The particular functional form of $D$ depends on the shape of the VDF [Equation (\ref{eq:vdf})]. For a bi-Maxwellian distribution given by Equation (\ref{eq:vdf}), the details can be found extensively in the literature~\citep[see e.g.][and references therein]{stix1992,vinas2000,Yoon2017}. The Vlasov-Maxwell linear analysis provides a robust framework to characterize the linear response of the media to the propagation of electromagnetic waves in different plasma environments. Considering Equation~(\ref{eq:dr}) as an eigen-value problem, $\delta\mathbf{E_k}$ corresponds to the fluctuating electric field (eigen-modes) of the plasma, and $\omega = \omega(\mathbf{k})$ corresponds to the solutions of the dispersion relation (eigen-frequencies) as a function of the wave-vector $\mathbf{k}$. Thus, the wave-normal angle $\theta$ between the mean field and the direction of propagation of the waves is given by $\cos(\theta) = \mathbf{k}\cdot\mathbf{B}_0/(k\, B_0)$, where $k = |\mathbf{k}|$ is the wave-number. 

Besides the solutions $\omega(\mathbf{k})$ of the dispersion relation, using Equation~(\ref{eq:dr}) and Maxwell's equations it is also possible to obtain information about the spectral properties of each eigen-mode $\delta\mathbf{E_k}(\omega, \mathbf{k})$, such as polarization $P = i \delta E_{kx}/\delta E_{ky}$; reduced magnetic helicity $\sigma_{mk} = \langle \delta \mathbf{A}_k\cdot \delta \mathbf{B}_k\rangle/|\delta\mathbf{B}_k|^2$, where $\delta \mathbf{A}$, and $\delta \mathbf{B}$, are the magnetic vector potential and magnetic field, respectively; magnetic compressibility $C_B = |\delta B_{kz}|/|\delta\mathbf{B}_k|$; and the ratio between parallel electric fluctuations and transverse magnetic fluctuations $\delta E_\parallel|/|\delta B_\bot| = |\delta E_{kz}|/(|\delta B_{kx}|^2+|\delta B_{ky}|)^{1/2} $, among other dispersion properties. Without loss of generality, for all these definitions we have assumed $\mathbf{k} = k_x \hat{x} + k_z \hat{z}$, such that $k = \sqrt{k^2_x+k^2_z}$ (see Appendix~\ref{appendix} for details).

The solutions of the dispersion relation, Equation~(\ref{eq:dr}), and the dispersion spectral properties of each wave-mode will vary depending on the particular value of the wave-normal angle, the composition of the plasma, and the characteristics and macroscopic parameters of the plasma. In the case of bi-Maxwellian VDFs, this dependence can be expressed in terms of a reduced set of dimensionless quantities such as the relative abundance $\eta_s = n_{0s}/n_0$ of each species $s$ with respect to the total density, the plasma beta $\beta_s = 8\pi n_s k_B T_s/B^2_0$ and the temperature anisotropy $\mu_s = T_{\bot s}/T_{\parallel s}$ of each species, the ratio between the local Alfvén and the speed of light $C_A = V_A/c$, and the angle of propagation of the waves. All these linear properties are well-known descriptors of plasma waves and have been widely used to characterize different wave-modes using theory, and are key information to identify them through observations~\citep[see e.g.][]{gary1986,gary1993,vinas2000,moya2015}. Each of these properties is important to discriminate between KAWs, EMIC or MSW modes~\citep{gary1986,salem2012,moya2015}, so that their combination will allow us to assess whether the presence of heavy ions such as O+ and He+ favor or not the existence and propagation of KAWs in the inner magnetosphere.

\section{Results: the effect of O+ and H\lowercase{e}+ ions}
\label{ref:results}

\setlength{\tabcolsep}{5pt}
\begin{deluxetable}{cccccc}
    \tablenum{1}
    \tablecaption{Linear Theory Calculation Parameters}
    %\tablewidth{0pt}
    \tablehead{$L$-shell & $B_0$ (nT) & $C_A = v_A/c$ & Species  & $\eta$ & $\beta$}
    \startdata
     &  & & H$^+$ & 0.40 & 0.068\\
    4 & 487.5 & $1.12\times10^{-2}$ & O$^+$ & 0.50 & 0.085\\
    &  & & He$^+$ & 0.10 & 0.017\\
    \hline
     &  & & H$^+$ & 0.50 & 0.323\\
    5 & 249.6 & $5.74\times10^{-3}$ & O$^+$ & 0.45 & 0.291\\
    &  & & He$^+$ & 0.05 & 0.032\\
    \hline
    &  & & H$^+$ & 0.58 & 1.120\\
    6 & 144.4 & $3.32\times10^{-3}$ & O$^+$ & 0.40 & 0.772\\
    &  & & He$^+$ & 0.02 & 0.039\\
    \enddata
    \tablecomments{Total electron density is $n_e$ =10 cm$^{-3}$, and $T=$10 keV for all species.}
    \label{tbl1}
\end{deluxetable}
To analyze the properties of KAW in the inner magnetosphere we consider realistic parameters of the ring current region, in which the plasma is composed by electrons, protons (H+ ions), O+ ions and He+ ions, with abundances that can vary significantly especially during geomagnetically disturbed periods. In particular, depending on the strength of a given geomagnetic storm, the abundance of O+ ions $\eta_{O+}$ can vary between 20\% and 80\%~\citep{hamilton1988,daglis1999}.  Moreover,~\citet{jahn2017} showed that O+ abundance can be parameterized as a function of $L$-shell and the Kp index, so that warm O+ can vary from $\eta_{O+}\sim$ 60\% at $L$=4, to $\eta_{O+}\sim$ 40\% at $L$=6 when Kp $\geq$ 4, and He+ abundance $\eta_{He+}$ does not exceed 10\% during geomagnetically active times for all $4<L<6$. To solve the dispersion relation with parameters representative of the ring current during geomagnetic storms we select the average abundances of H+, O+ and He+ ions reported by~\citet{jahn2017} at $L=4$, $L=5$, and $L=6$ during geomagnetic storms. We consider a quasi-neutral plasma in which the total ion density $n_0$ is equal to the total electron density $n_e$, and then $\eta_{H+} \, + \, \eta_{O+} \, + \, \eta_{He+} = 1$. 
\begin{figure}[ht]
	\centering
		\includegraphics[width=.99\linewidth]{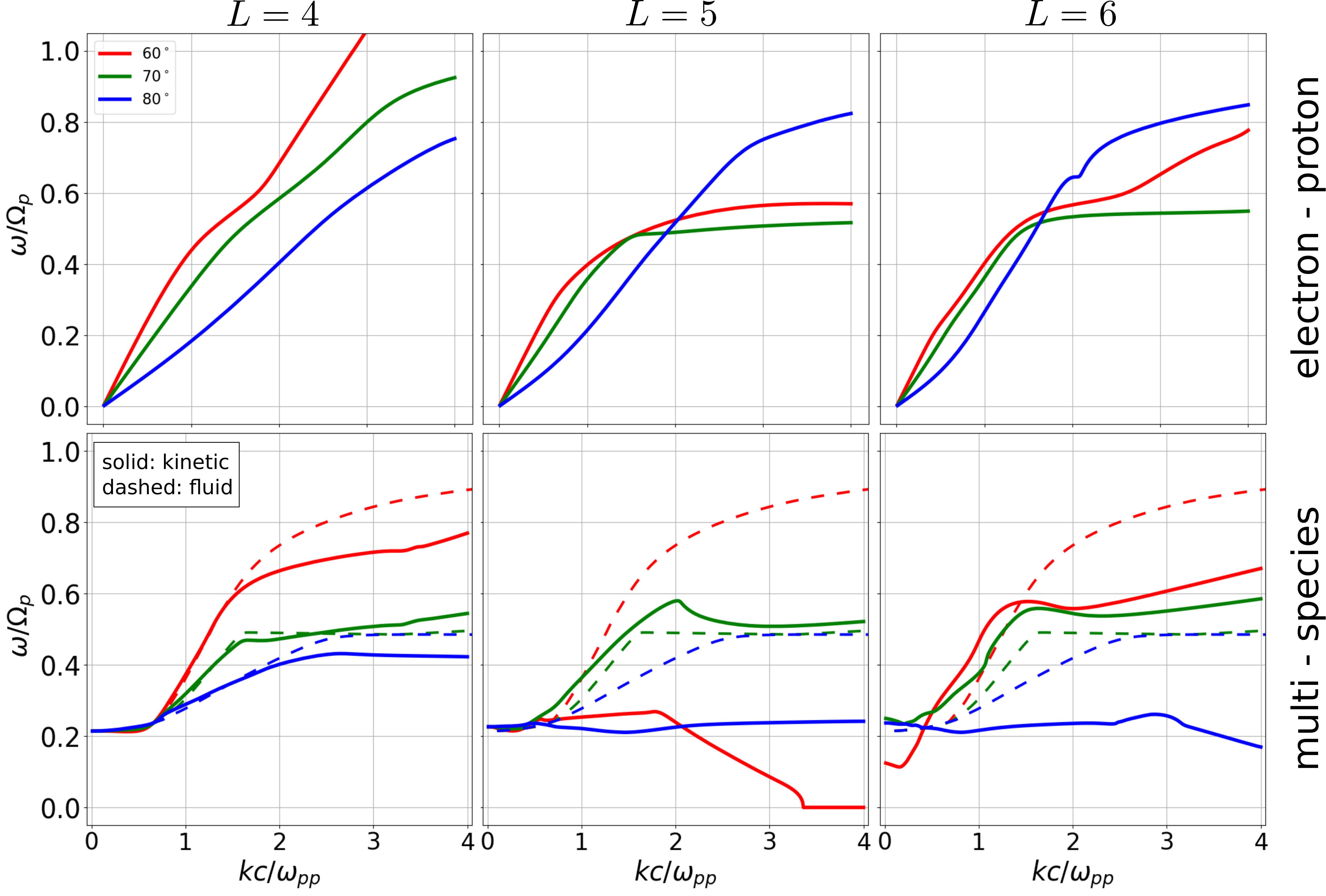}
	\caption{Electron-proton (top) and multi-species plasma (bottom) dispersion relation of Alfvénic solutions for wave-normal angle $\theta$ = 60$^{\circ}$ (red), 70$^{\circ}$ (green), and 80$^{\circ}$ (blue), considering parameters shown in Table~\ref{tbl1} for $L$ = 4 (left), $L$ = 5 (center), and $L$ = 6 (right). Frequency and wave-number are expressed in units of the proton gyro-frequency and the proton inertial length, respectively. In bottom panels solid and dashed lines represent the kinetic and fluid dispersion relation, respectively.
	}
	\label{FIG:1}
\end{figure}

The beta parameter of each species at each $L$ shell is calculated considering (for simplicity) a non-drifting ($U_s =0$), isotropic $(\mu_s=1)$, and isothermal ($T_s = T$) plasma with a temperature $T=10$~keV typical for the ring current region~\citep{kamide}, and total electron number density $n_e=10$~cm$^{-3}$. We also assume a dipolar field so that at each $L$-shell $B_0\sim B_D/L^3$, and then $\beta_s \propto n_s T L^6/B^2_D$, where $B_D = 3.12\times 10^4$~nT is the average strength of the magnetic field at the magnetic equator at $L=1$. With this information all the necessary dimensionless parameters are determined and shown in Table~\ref{tbl1}. The only free parameter remaining is the wave-normal angle. It is important to mention that, depending on the position of the plasmapause, the warm or hot populations may not dominate the composition of the plasma at $4<L<6$, so that our selection of parameters may not be appropriate in such cases. However, during geomagnetic active times, when plasmaspheric plumes transport the cold plasma towards the magnetopause~\citep{foster2014} the plasmasphere is eroded and confined to lower $L$-shells, so it is reasonable to consider $L \geq 4$ to be outside the plasmasphere. For example, in~\citep{OBrien2003} it is shown that empirical models predict a plasmapause at $L\sim 4.18$ for Kp = 4, and at $L < 4$ for Kp $>$ 5. Therefore, for geomagnetically active times, our model of the region of interest is a reasonable choice. Furthermore, as the important quantity that seems to control the shift from EMIC waves to KAWs is plasma beta~\citep{gary1993,Gary2004}, and considering that plasma beta depends on a particular combination of density, temperature and magnetic field, other values of electron density and temperature combined with a different model for the magnetic field strength could lead to the same beta values shown in Table~\ref{tbl1}, and therefore  the same results from the kinetic theory analysis~\citep{moya2015}. 

\begin{figure}[ht]
	\centering
		\includegraphics[width=.99\linewidth]{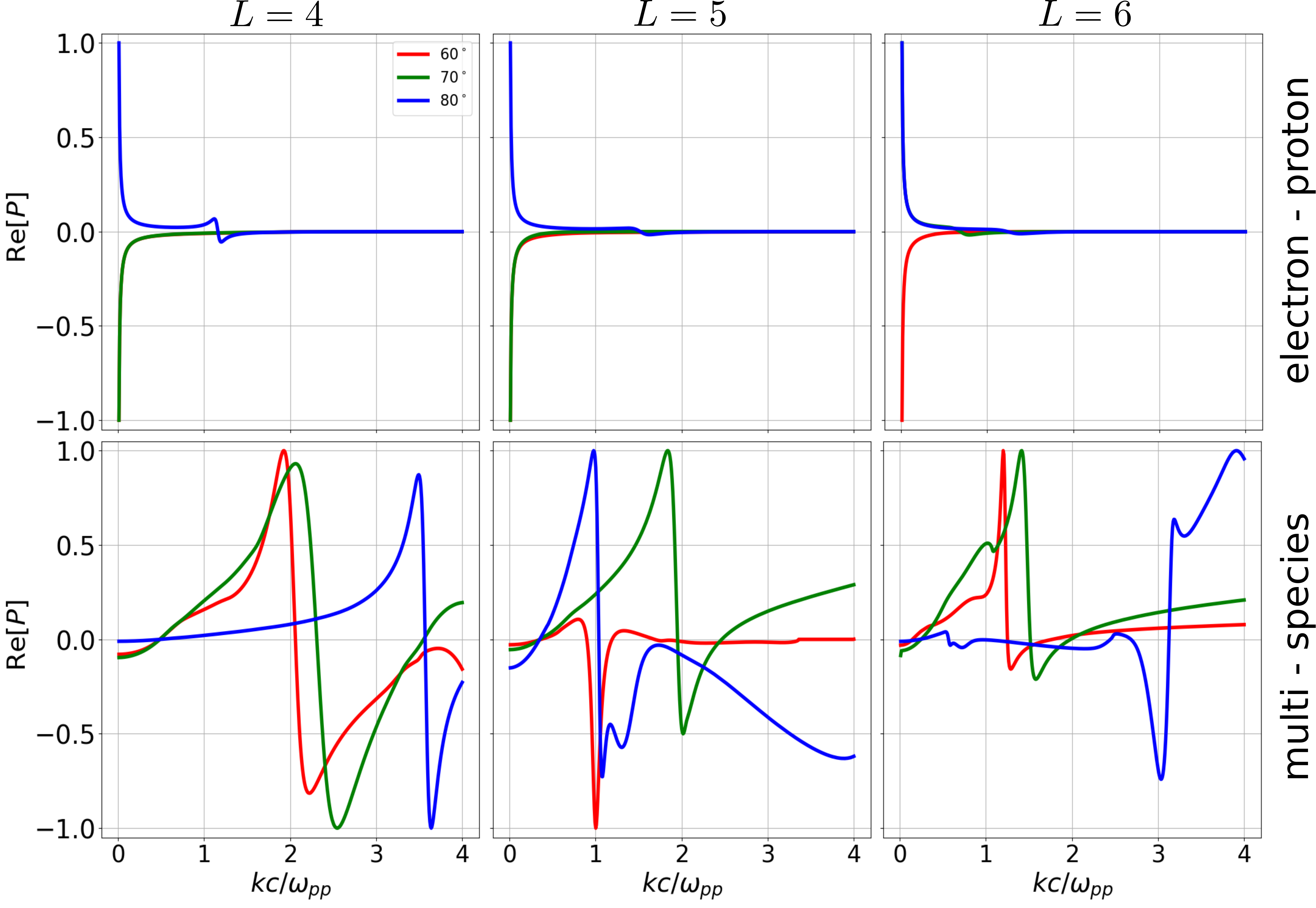}
	\caption{Polarization spectra for the Alfvénic modes shown in Fig.~\ref{FIG:1},for electron-proton (top) and multi-species plasma (bottom), for $L$ = 4 (left), $L$ = 5 (center), and $L$ = 6 (right). Red, green and blue curves represent wave-normal angle of 60$^{\circ}$, 70$^{\circ}$, and 80$^{\circ}$, respectively. considering parameters shown in Table~\ref{tbl1}. Each polarization curve is normalized to its maximum value, and the wave-number expressed in units of the proton inertial length.
	}
	\label{FIG:2}
\end{figure}

To study how the presence of both O+ and He+ ions, together with protons and electrons, influence the existence of KAWs at each $L$-shell shown in Table~\ref{tbl1}, we numerically solve the exact dispersion relation considering wave-normal angles of $\theta = 60^{\circ}$, $70^{\circ}$, and $80^{\circ}$, as these propagation angles should be the most suitable for a KAW solution. We then compare the dispersion relation and dispersion properties spectra (polarization, helicity, magnetic compressibility, and parallel electric field to transverse magnetic field ratio) obtained for the multi-species magnetospheric plasma with an electron-proton plasma of the same density and temperature for each considered $L$ and $\theta$. The calculation and comparison was done without the use of any approximation to the elements of the dispersion tensor something possible as we use our own kinetic dispersion solver written in Python \citep{moya2021}. Fig.~\ref{FIG:1} shows a comparison between the dispersion relation obtained considering a pure electron-proton plasma (top panels) and the magnetospheric multi-species case (bottom panels), for wave-normal angle $\theta$ = 60$^{\circ}$ (red), 70$^{\circ}$ (green), and 80$^{\circ}$ (blue), and the three $L$-shells shown in Table~\ref{tbl1}. Further, in the multi-species case, from the three Alfvénic solutions that exist in a plasma composed by three different ion species~\citep[see e.g.][and references therein]{Saikin2015,Blum2017}, here we only show the results associated with the H+ band, as this frequency band is the most similar with the Alfvén branch in an electron-proton plasma compared to the O+ or He+ bands. Moreover, for comparison purposes, in the multi-species case we have also included the solutions of the cold multi fluid dispersion relation~\citep{stix1992,fitzpatrick} (dashed lines in Fig.~\ref{FIG:1}). From this figure we can clearly appreciate the effect of plasma beta and the propagation angle. Even for the case with the smaller beta values ($L$ = 4; left panel) the figure shows how the dispersion relation does not follow the typical Alfvénic solutions obtained with the cold multi fluid plasma approximation. When heavy ions are considered while computing the susceptibility of the media, the departure from the multi fluid dispersion relation (dashed lines in Fig.~\ref{FIG:1}) is evident. In summary, these ions can clearly affect the dispersion tensor and dispersive properties of the plasma, 

\begin{figure}[ht]
	\centering
		\includegraphics[width=1.0\linewidth]{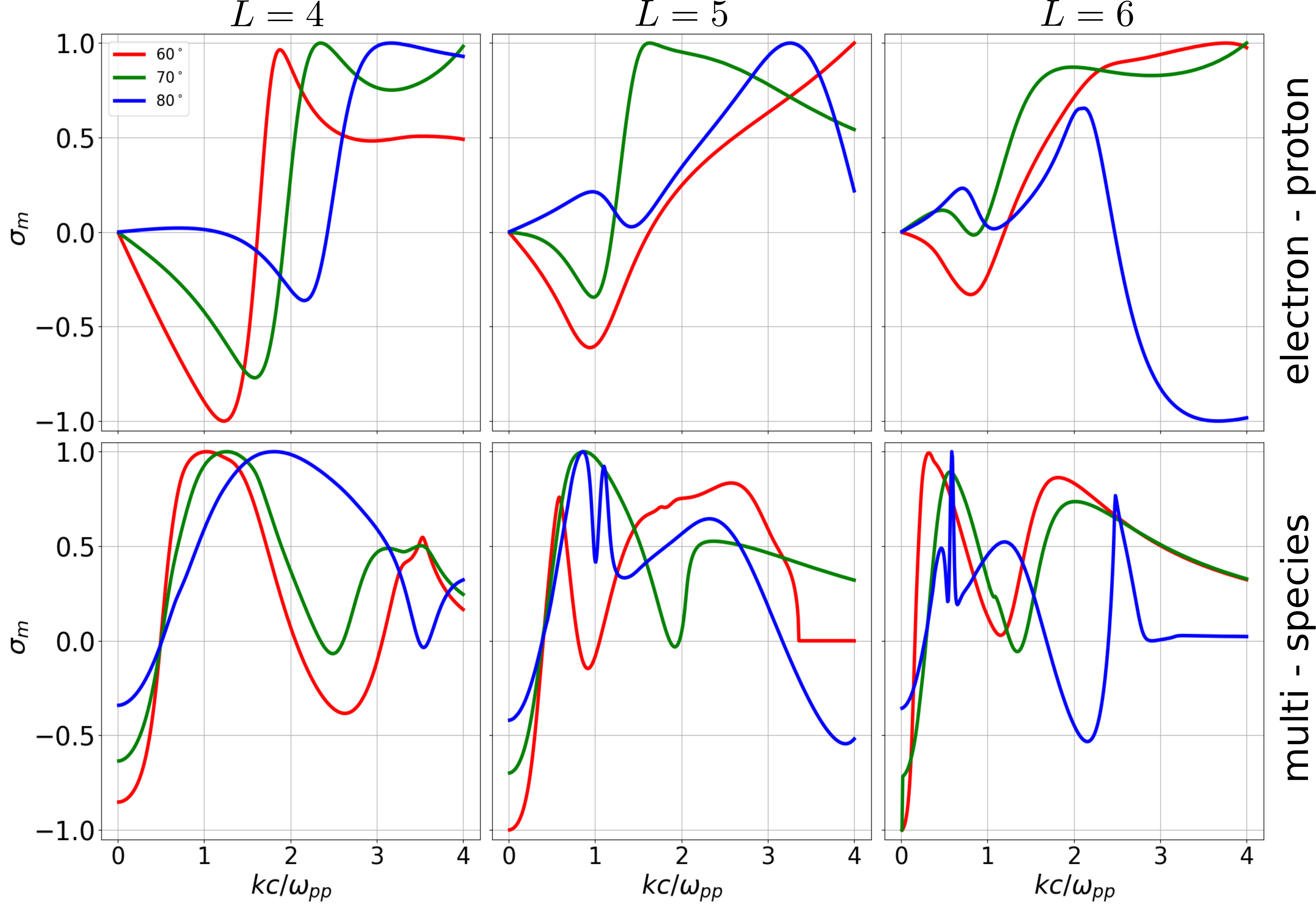}
	\caption{Electron-proton (top) and multi-species plasma (bottom) magnetic helicity spectra for wave-normal angle $\theta$ = 60$^{\circ}$ (red), 70$^{\circ}$ (green), and 80$^{\circ}$ (blue), considering parameters shown in Table 1 for $L$ = 4 (left), $L$ = 5 (center), and $L$ = 6 (right). Wave-numbers are expressed in units of the proton inertial length.}
	\label{FIG:3}
\end{figure}
To analyze the effect of the heavier ions on the existence of KAWs we need to search for the signature characteristics of KAWs in the dispersion properties of the Alfvénic modes shown in Fig.~\ref{FIG:1}; namely: positive polarization and helicity, large magnetic compressibility, and parallel-electric to transverse-magnetic fluctuations ratio. For each of the dispersion relation solutions presented in Fig.~\ref{FIG:1}, in Fig.~\ref{FIG:2} we show the polarization spectra (normalized to its maximum value). From Fig.~\ref{FIG:2} we can see that at all $L$-shells in the electron-proton plasma case (top panels) the polarization is positive only for $\theta$ = 80$^{\circ}$ and $k c / \omega_{pp} \sim$ 1. In contrast, bottom panels in Fig.~\ref{FIG:2} show that the multi-species plasma allows Alfvénic solutions with positive polarization for all three considered wave-normal angles, consistent with KAW modes, and a wider wave-number range. In general this region moves from smaller to larger wave-number values (larger to smaller scales) as the wave-normal angle increases and goes up to $k c / \omega_{pp} \sim$ 4 for $\theta$ = 80$^{\circ}$ at $L$ = 6, but the range is relatively smaller when compared to the range at $L$ = 4 and 5. Note that for this last case the solution at 80$^{\circ}$ exhibits positive polarization only for $k c / \omega_{pp} > 3$. Thus, even though the abundance (and therefore the influence) of heavy ions decreases with increasing $L$-shell, as the magnetic field is weaker further from the Earth, the plasma beta of all species increases and the solutions with positive polarization (compatible with KAW modes) still exist, even though the range in wave-number with right-handed solutions is not the same for all $L$-shells. It is important to note that the first principles reasons for the changes in polarization in different $k$ intervals as a function of composition, wave-normal angle and plasma beta must be related with the properties of the dispersion tensor, and subsequently its eigen-modes (the electric field of the waves). That being said, to the best of our knowledge the exact reason for this phenomena is still unknown. Indeed, even in the electron-proton case the answer to these questions is not trivial at all. We may speculate about possible changes in sign, magnitude,  and/or topology of some of the elements of the dispersion tensor (see Appedix \ref{appendix}) in the complex plane depending on the values of $k_\parallel = k \cos(\theta)$ and $k_\bot = k \sin(\theta)$. However, the exact answer on why plasma waves transit from EMIC to KAW as plasma beta and wave normal angle increase is beyond the scope of this study.

\begin{figure}[ht]
	\centering
		\includegraphics[width=1.0\linewidth]{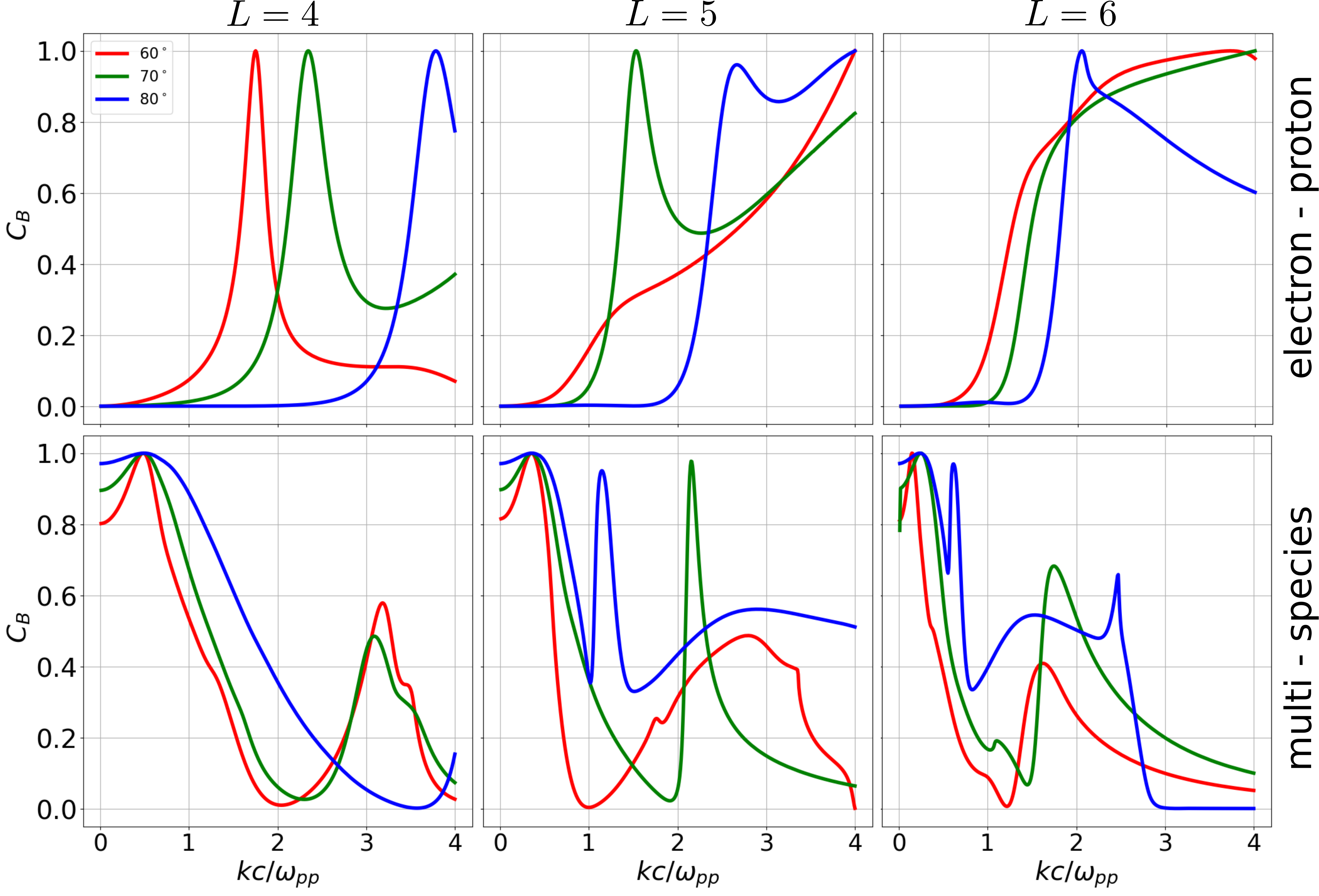}
	\caption{Electron-proton (top) and multi-species plasma (bottom) magnetic compressibility spectra for wave-normal angle $\theta$ = 60$^{\circ}$ (red), 70$^{\circ}$ (green), and 80$^{\circ}$ (blue), considering parameters shown in Table 1 for $L$ = 4 (left), $L$ = 5 (center), and $L$ = 6 (right). Wave-numbers are expressed in units of the proton inertial length.}
	\label{FIG:4}
\end{figure}

The differences in the dispersion properties between the electron-proton and multi-species plasmas can also be seen in the magnetic helicity and magnetic compressibility spectra, in which, regardless of the wave-normal angles or $L$-shell, the presence of O+ and He+ ions introduces non-trivial changes on the helicity and magnetic compressibility. Regarding the helicity, and consistent with the polarization, Fig.~\ref{FIG:3} shows that in multi-species plasmas, the wave modes have positive and larger helicity at smaller wave-number values compared with an electron-proton plasma at all considered wave-normal angles and at all different $L$-shell. Furthermore, comparing with Fig.~\ref{FIG:2} we can see that in all cases both polarization and helicity are positive in the same wavenumber range, as expected for KAW modes. Moreover, regarding the magnetic compressibility spectra, Fig.~\ref{FIG:4} shows that, in contrast with the electron-proton case, the multi-ion plasma allows solutions with large magnetic compressibility $C_B \gtrapprox 0.5$ even for $kc/\omega_{pp} = 1$, a wavenumber range in which helicity and polarization are positive as well, which is not possible in an electron-proton plasma at all considered wave-normal angles.

\begin{figure}[ht]
	\centering
		\includegraphics[width=.99\linewidth]{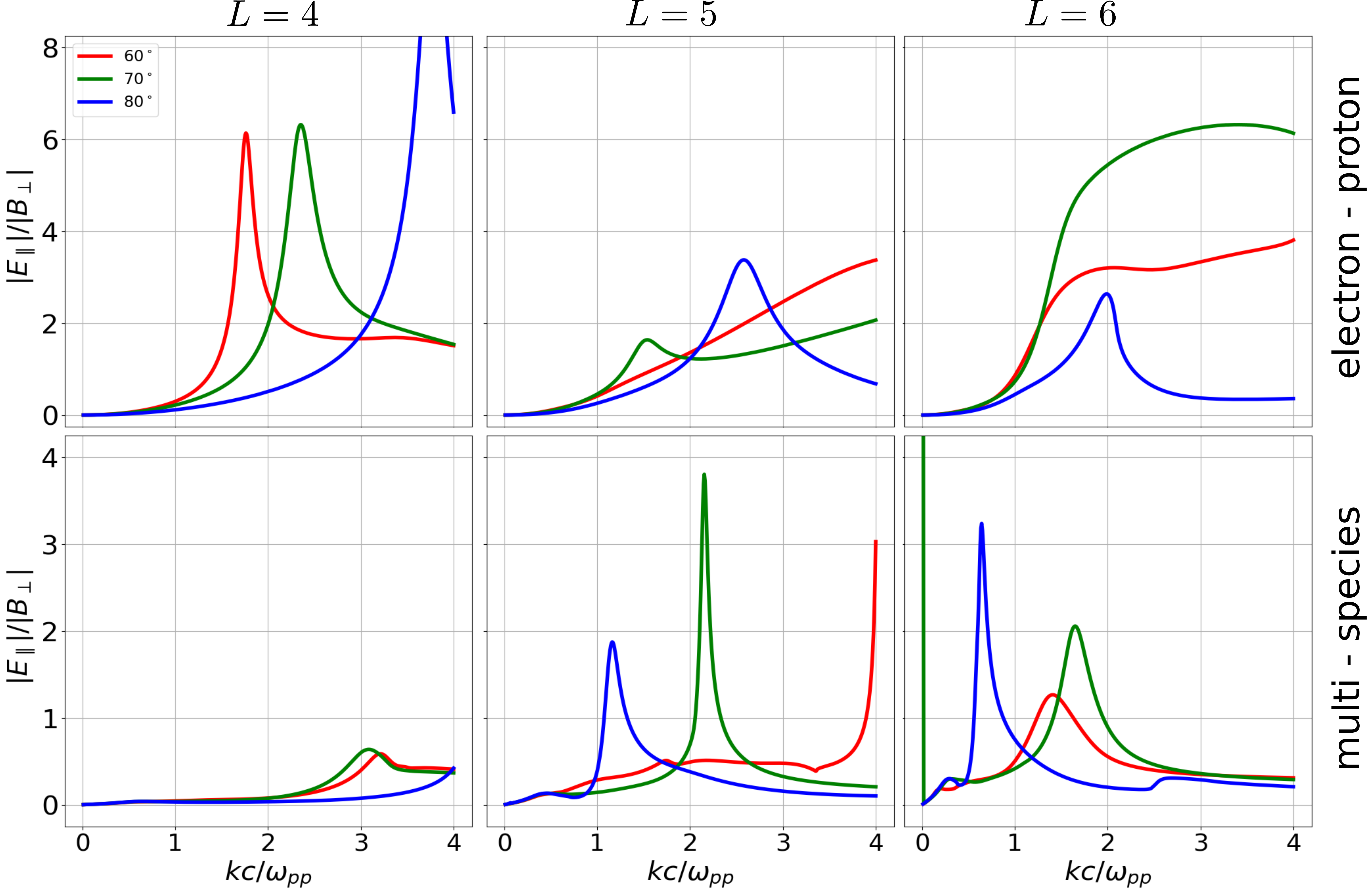}
	\caption{Parallel-electric-field to transverse-magnetic-field ratio spectra, $\delta E_\parallel|/|\delta B_\bot|$, for electron-proton (top) and multi-species magnetospheric plasma (bottom), considering the same wave-normal angles and $L$-shells as in all previous figures. In all panels $\delta E_\parallel|/|\delta B_\bot|$ is expressed in units of the local Alfvén speed $v_A$, and the wave-number is normalized to the proton inertial length.}
	\label{Fig:5}
\end{figure}

Thus, the multi-species plasma can allow the propagation of magnetically compressive Alfvénic waves with positive magnetic helicity, at wave-numbers not accessible if the heavier ions were not present. Finally, even though compressive Alfvénic modes with positive polarization, and also positive magnetic helicity are good candidates for KAWs and clearly not for EMIC waves, MSW can also share these properties. As already mentioned, a key signature to distinguish between MSW and KAWs is the $\delta E_\parallel/\delta B_\bot$ spectrum~\citep{salem2012,chaston2014,moya2015}. Fig.~\ref{Fig:5} shows these spectra, in units of the local Alfvén speed, for all considered wave-normal angles and $L$ shells for both cases: the electron-proton (top panels) and the multi-species (bottom panels) plasma. From Fig.~\ref{Fig:5} we can see that for the electron-proton case, only for $L$ = 4 the $\delta E_\parallel/\delta B_\bot$ spectrum has a clear peak for the three considered wave-normal angles. For $L$ = 5 and $L$ = 6, only the solution at 80$^{\circ}$ exhibits a $\delta E_\parallel/\delta B_\bot$ spectrum compatible with KAWs. In contrast, when He+ and O+ are considered in the dispersion relation the situation is the opposite. for $L$ = 4 the $\delta E_\parallel/\delta B_\bot$ is less than one at all wave-normal angles and wave-numbers showing that in this case the Alfvénic modes are not KAWs. Therefore, even though in this case where the polarization is positive (see Fig.~\ref{FIG:2}), as the plasma beta values are rather small, the $k_{\perp} \rho_i \sim 1$ condition is not fulfilled and the KAW modes are not present (in this case the solutions correspond to MSW modes). However, for larger $L$ shells all the $\delta E_\parallel/\delta B_\bot$ exhibit a peak larger than one. This shows that the obtained solutions at $L$ = 5 and $L$ = 6 are indeed Kinetic Alfvén Wave modes, and that the existence of the KAWs is only possible due to the presence of the heavier magnetospheric ions. 

\section{Discussion and Conclusions}
\label{ref:conclusions}

By using the Vlasov-Maxwell linear theory of plasma waves we have obtained the dispersion relation and analyzed the dispersion properties of Alfvénic modes in two different scenarios: an electron-proton plasma and a magnetospheric multi-species plasma composed by electrons, protons, He+ and O+ ions, with plasma parameters motivated by in situ observations in the inner magnetosphere. We analyzed the changes introduced in the dispersion relation and dispersion properties such as polarization, magnetic helicity, magnetic compressibility, and electric and magnetic field perturbations due to the inclusion of heavier ions (in combination with protons and electrons). We compared the possibility of obtaining Kinetic Alfvén Wave mode solutions in both scenarios, as in a multi-species plasma, such as the inner magnetosphere, the properties of the KAW should not only depend on the parameters of proton and electron distributions, but also on the relative abundances of heavier ions and their properties. As shown in Table~\ref{tbl1}, and depending on the $L$-shell, thanks to the shrinking of the plasmapause to $L < 4$ during geomagnetic active intervals, it is possible that the dominant ion species in the inner magnetosphere can correspond to O+ and not to protons. 

In the electron-proton case, our results show that considering temperatures and number density values relevant for the ring current region outside the plasmasphere, for $kc/\omega_{pp} \leq 4$ the presence of Kinetic Alfvén Waves is not possible at wave-normal angles $\theta \leq$  70$^{\circ}$. In contrast, the presence of He+ and O+ ions introduces significant changes on the dispersion properties of Alfvénic modes, allowing the existence of KAW mode solutions at $L \geq$ 5, at even $\theta = $  60$^{\circ}$, which is consistent with observations~\citep[see e.g.][]{moya2015}. In addition, compared to the electron-proton case, the KAW solutions are possible in wider wave-number range, and in general the range shifts from smaller to larger wave-number values (lager to smaller scales) as the wave-normal angle increases. 

It is important to emphasize that in the inner magnetosphere, KAWs have been predominately observed during geomagnetically disturbed time intervals, either during substorms in association with particle injections from the magnetospheric tail~\citep{wygant2002}, or during geomagnetic storms~\citep{chaston2014,Chaston2015a,Chaston2015b,Chaston2020,moya2015}, during roughly the same time period of time in which the abundance of heavy ions increase~\citep{jahn2017}. This is consistent with our own findings that suggest that in the inner magnetosphere, the conditions necessary for the occurrence of KAWs in multi-species plasmas are those that can also be found during geomagnetic storms. 

In summary, we have shown that for plasma parameters relevant for the ring current outside of the plasmasphere during geomagnetic storms, the propagation of Kinetic Alfvén Waves is highly determined by the presence of O+ and He+. Thus, as wave-particle interactions with KAW correspond to a possible channel mechanism for the energy transfer from ions to electrons, our results suggest that magnetospheric ions may play an important role on the energization of sub-ionic and electronic spatial scales and the acceleration of plasma particles, especially during intense geomagnetic storms in which O+ ions can dominate the plasma composition in the inner magnetosphere. However, in order to quantify the relevance of this processes it is necessary to increase the scope of the study by analyzing the complex frequency (growth/damping rate) of the waves, and also include non-linear effects necessary to properly assess the energy transfer between KAWs and plasma particles. We expect this theoretical analysis motivated by observations, to provide evidence that for many types of studies based on particle or fluid simulations, and linear, quasi-linear or non-linear models, the inclusion of heavy ions should be customary for realistic modeling of plasma phenomena in the inner magnetosphere, or other space environments where heavy ions contribute with a substantial portion of the plasma density, such as planetary magnetospheres and comet plasma tails.

\acknowledgements

\noindent We are grateful for the support of ANID, Chile through FONDECYT grant No. 1191351 (PSM) and National Doctoral Scholarships Nos. 21181965(BZQ), 21182002 (IGM), and SIA SA772100112 (VAP). Dispersion analysis output files are available at https://doi.org/10.5281/zenodo.3900556.

\appendix

\section{Dispersion tensor and dispersion properties}
\label{appendix}

Here we briefly present basic expressions for the dispersion tensor and dispersion properties of electromagnetic waves propagating through a 
magnetized, colissionless, and uniform bi-Maxwellian plasma with distribution functions given by Eq.\eqref{eq:vdf}.

\subsection{Dispersion tensor}
    Considering the background magnetic field as $\mathbf{B_0} = B_0\hat{z}$, and wave-vectors for parallel propagating waves given by $\mathbf{k} = k_\perp \hat x + k_\parallel \hat{z}$ (where $k_\parallel = k \cos(\theta)$ and $k_\bot = k \sin(\theta)$, with $\theta$ the wave-normal angle), the dispersion tensor $D(\omega, \mathbf{k})$ can be written in the following form~\citep{stix1992,vinas2000,moya2021}:
        \begin{equation}
            \label{eq:lambda}
            D(\omega, \mathbf{k}) = 
            \begin{pmatrix}
                D_{xx}& D_{xy}  & D_{xz} \\ 
                -D_{xy} & D_{yy} & D_{yz} \\ 
                D_{xz}&  -D_{yz} & D_{zz}
            \end{pmatrix} \,,
        \end{equation}
        where each element is given by
       
        \begin{equation}
            \label{eq:dxx}
            D_{xx} = 1 - \dfrac{k_{\parallel}^2c^2}{\omega^2} + 2\sum_s \sum_{n=-\infty}^{\infty} \dfrac{\omega_{ps}^2}{\omega^2} \Lambda_n(\lambda_s) \left(\dfrac{n^2 \Omega_s^2}{k_{\bot}^2\alpha_{\bot s}^2}\right) [(\mu_s - 1) +\mu_sZ(\xi_{ns})\bar{\xi}_{ns}]\,,    
        \end{equation}
    
        \begin{equation}
            D_{xy} = i\sum_s \sum_{n=-\infty}^{\infty} \dfrac{\omega_{ps}^2}{\omega^2}\mu_s n \Lambda'(\lambda_s) Z(\xi_{ns}) \bar{\xi}_{ns}\,,    
        \end{equation}
    
        \begin{equation}
            D_{xz} = \dfrac{k_{\parallel}k_{\bot}c^2}{\omega^2} + 2\sum_s\sum_{n=-\infty}^{\infty} \dfrac{\omega_{ps}^2}{\omega^2} \Lambda_n(\lambda_s)\left(\dfrac{n\Omega_s}{k_{\bot}\alpha_{\parallel s}}\right)\left[\left(\mu_s^{-1} - 1\right) \left(\dfrac{n\Omega_s}{k_{\parallel}\alpha_{\parallel s}}\right) + y_{ns} Z(\xi_{ns}) \bar{\xi}_{ns}\right]\,,    
        \end{equation}
    
        \begin{equation}
            D_{yy} = 1 - \dfrac{k^2c^2}{\omega^2} + 2\sum_s\sum_{n=-\infty}^{\infty} \dfrac{\omega_{ps}^2}{\omega^2} \left(\dfrac{\Omega_s^2}{k_{\bot}^2\alpha_{\bot s}^2}\right) (n^2\Lambda_n(\lambda_s)-2\lambda_s^2\Lambda'_n(\lambda_s)) \left[(\mu_s - 1) + \mu_s Z(\xi_{ns}) \bar{\xi}_{ns}\right]\,,   
        \end{equation}
    
        \begin{equation}
            D_{yz} = -2i\sum_s\sum_{n=-\infty}^{\infty} \dfrac{\omega_{ps}^2}{\omega^2} \Lambda'_n(\lambda_s)\left(\dfrac{\lambda_s\Omega_s}{k_{\bot}\alpha_{\parallel s}}\right)y_{ns} Z(\xi_{ns}) \bar{\xi}_{ns}\,,
        \end{equation}
    
        \begin{equation}
            D_{zz} = 1 - \dfrac{k_{\bot}^2c^2}{\omega^2} + 2\sum_s\sum_{n=-\infty}^{\infty} \dfrac{\omega_{ps}^2}{\omega^2} \Lambda_n(\lambda_s)\left[\left(\dfrac{\omega}{k_{\parallel} \alpha_{\parallel s}}\right)^2 + \left(1 - \mu_s^{-1}\right)\left(\dfrac{n\Omega_s}{k_{\parallel}\alpha_{\parallel s}}\right)^2 + y_{ns}^2 Z(\xi_{ns}) \bar{\xi}_{ns}\right]\,.   
        \end{equation}
        Here $\omega_{ps} = 4\pi \,n_{0s}\, q^2/m_s,$ and $\Omega_s = q_s B_0/m_s c$ are the plasma and gyro-frequency of the species $s$, and 
        $\mu_s = T_{\bot s}/T_{\parallel s}$ is its temperature anisotropy. In addition, we have defined the following quantities
        
        \begin{equation}
            \xi_{ns} = \dfrac{\omega - n\Omega_s - k_{\parallel}U_{\parallel s} }{k_{\parallel}\alpha_{\parallel s}}\,, \quad
            \bar{\xi}_{ns} = \dfrac{\omega - n\Omega_s(1 - \mu_s^{-1}) - k_{\parallel}U_{\parallel s}}{k_{\parallel}\alpha_{\parallel s}}\,,\quad
            y_{ns} = \dfrac{\omega - n\Omega_s }{k_{\parallel}\alpha_{\parallel s}}\,,
        \end{equation}
        where $I_n$ represents the $n$-th modified Bessel of the first kind,

        \begin{equation}
            \label{eq:bessel}
            \lambda_s = \dfrac{1}{2}\dfrac{k_{\bot}^2 \alpha_{\bot s}^2}{\Omega_s^2}\,,\quad
            \rm{and} \quad \Lambda_n(\lambda_s) = e^{-\lambda_s}I_n(\lambda_s)\,.
        \end{equation}

\subsection{Kinetic dispersion properties}
    
 As the dispersion relation implies that $|D(\omega,\mathbf{k})| = 0$, it is not possible to obtain the three components of the fluctuating electric field $\delta \mathbf{E_k} = \delta E_{kx}\hat x + \delta E_{ky}\hat y  + \delta E_{kz}\hat z$ independently. Here we impose the normalization $|\delta E_{kx}|^2+|\delta E_{ky}|^2+|\delta E_{kz}|^2 = 1$, and obtain:
\begin{eqnarray}
\label{eq:ey}
\dfrac{\delta E_{ky}}{\delta E_{kx}} &= \dfrac{1}{2}\left(\dfrac{D_{yz}D_{xx}-D_{yx}D_{xz}}{D_{yy}D_{xz}-D_{yz}D_{xy}}+\dfrac{D_{xx}D_{zz}-D_{zx}D_{xz}}{D_{zy}D_{xz}-D_{zz}D_{xy}}\right)\\
\label{eq:ex}
\dfrac{\delta E_{kz}}{\delta E_{kx}} &= \dfrac{1}{2}\left(\dfrac{D_{xx}D_{yy}-D_{xy}D_{yx}}{D_{yy}D_{xz}-D_{yz}D_{xy}}+\dfrac{D_{xx}D_{zy}-D_{xy}D_{zx}}{D_{zy}D_{xz}-D_{zz}D_{xy}}\right)\,.
\end{eqnarray}

In addition, the fluctuating magnetic field is given by Faraday's law:
\begin{equation}
\label{eq:b}
    \delta \mathbf{B_k} = \dfrac{c\,\mathbf{k}}{\omega}\times \delta \mathbf{E_k}\,,
\end{equation}
so that
\begin{eqnarray}
\label{eq:bx}
\delta{B_{kx}} &=& -\dfrac{c\,k_{\parallel}}{\omega} \,\delta E_{ky}\,,\\
\label{eq:by}
\delta{B_{ky}} &=& 
\dfrac{c\,k_{\parallel}}{\omega} \,\delta E_{kx} - \dfrac{c\,k_{\bot}}{\omega} \,\delta E_{kz}\,,\\
\label{eq:bz}
\delta{B_{kz}} &=& \dfrac{c\,k_{\bot}}{\omega}\, \delta E_{ky}\,.
\end{eqnarray}
Thus, using Eqs.~\eqref{eq:ey}-\eqref{eq:bz} we can define all relevant dispersion properties such as polarization, reduced magnetic helicity, magnetic compressibility, and parallel electric field-to-transverse magnetic field ratio in terms of the fluctuating fields. Following~\citet{gary1993}, here we provide a list of all these quantities. Namely:

\begin{itemize}
\item Polarization
\begin{equation}
 \label{eq:polarization}
P = i\dfrac{\delta E_{\mathbf{kx}}}{\delta E_{\mathbf{ky}}} = i\left(\dfrac{D_{xz}D_{yy}-D_{xy}D_{yz}}{D_{xx}D_{yz}- D_{xz}D_{yx}}\right)\,.
\end{equation}

\item Reduced magnetic helicity
\begin{equation}
    \label{eq:helicity}
    \sigma_{mk} = \dfrac{\langle \delta \mathbf{A_k}\cdot \delta \mathbf{B_k} \rangle}{|\delta \mathbf{B_k}|^2} = \dfrac{2\,\rm{Im\left[\delta E_{ky}\delta E^{*}_{kx}\cos(\theta)+\delta E_{kz}\delta E^{*}_{ky}\sin(\theta)\right]}}{|\delta E_{kx}|^2\cos^2(\theta)+|\delta E_{ky}|^2+|\delta E_{kz}|^2\sin^2(\theta)-R_{xz}(\theta)}\,,
\end{equation}
where $\mathbf{A_k}$ is the fluctuating vector potential, given by $\mathbf{B_k} = i\mathbf{k}\times\mathbf{A_k}$, and we have defined $R_{xz}(\theta) = \rm{Re}\left[\delta E_{kx}\,\delta E^{*}_{kz}\sin(2\theta)\right]$.

\item Magnetic compressibility
\begin{equation}
    \label{eq:comp}
    C_B = \dfrac{|\delta B_{kz}|}{|\delta\mathbf{B_k}|} = \dfrac{|\delta E_{ky}|\sin(\theta)}
    {\left[]|\delta E_{kx}|^2\cos^2(\theta)+|\delta E_{ky}|^2
    +|\delta E_{kz}|^2\sin^2(\theta)-R_{xz}(\theta)               \right]^{1/2}}\,.
\end{equation}

\item Ratio between parallel electric field and transverse magnetic field
\begin{equation}
\label{eq:epalbper}
\dfrac{\delta E_\parallel}{|\delta B_\bot|} = \frac{|\delta E_{kx}|}{\left(|\delta B_{ky}|^2+|\delta B_{kz}|^2\right)^{1/2}} =  \dfrac{|\omega|}{ck}\,\dfrac{|\delta E_{kz}|}{\left[(|\delta E_{kx}|^2+|\delta E_{ky}|^2)\cos^2(\theta))+|\delta E_{kz}|^2\sin^2(\theta)-R_{xz}(\theta)\right]^{1/2}}
\end{equation}

\end{itemize}

%\bibliographystyle{IEEEtran}
%{\small
%\bibliography{poster}}

\bibliography{KAW}
\bibliographystyle{aasjournal}

\end{document}